\documentclass[
    ,final            % use final for the camera ready runs
  ]
  {aipproc}

\layoutstyle{6x9}

\begin{document}

\title{Analysis of Two Eclipsing Hot Subdwarf Binaries with a Low Mass Stellar and a Brown Dwarf Companion}


\author{Veronika Schaffenroth}{
  address={Dr.~Karl Remeis-Observatory \& ECAP, Astronomical Institute,
Friedrich-Alexander University Erlangen-Nuremberg, Sternwartstr.~7, 96049 Bamberg, Germany}
}
\author{Stephan Geier}{
  address={Dr.~Karl Remeis-Observatory \& ECAP, Astronomical Institute,
Friedrich-Alexander University Erlangen-Nuremberg, Sternwartstr.~7, 96049 Bamberg, Germany}
}
\author{Ulrich Heber}{
  address={Dr.~Karl Remeis-Observatory \& ECAP, Astronomical Institute,
Friedrich-Alexander University Erlangen-Nuremberg, Sternwartstr.~7, 96049 Bamberg, Germany}
}\newpage
\author{Horst Drechsel}{
  address={Dr.~Karl Remeis-Observatory \& ECAP, Astronomical Institute,
Friedrich-Alexander University Erlangen-Nuremberg, Sternwartstr.~7, 96049 Bamberg, Germany}
}
\author{Roy H.\ \O stensen}{
  address={Institute of Astronomy, K.U.\ Leuven, Celestijnenlaan 200D, 3001 Heverlee, Belgium}
}
\author{Pierre F.\ L.\ Maxted}{
  address={Astrophysics Group, Keele University, Staffordshire, ST5 5BG, UK}
}
\author{Thomas Kupfer}{
  address={Dr.~Karl Remeis-Observatory \& ECAP, Astronomical Institute,
Friedrich-Alexander University Erlangen-Nuremberg, Sternwartstr.~7, 96049 Bamberg, Germany}
}
\author{Brad~N.~Barlow}{
  address={Department of Physics and Astronomy, University of North Carolina, Chapel Hill, NC~27599-3255, USA}
}
\author{the MUCHFUSS collaboration}{address={}}

\begin{abstract}
The formation of hot subdwarf stars (sdBs), which are core helium-burning stars located on the extended horizontal branch, is still not understood. Many of the known hot subdwarf stars reside in close binary systems with short orbital periods between a few hours and a few days with either M star or white dwarf companions. Common envelope ejection is the most probable formation channel. Among these, eclipsing systems are of special importance because it is possible to constrain the parameters of both components tightly by combining spectroscopic and light curve analyses. We report the discovery of two eclipsing binaries with a brown dwarf ($<0.07\,M_{\rm \odot}$ ) and a $0.15\,M_{\rm \odot}$ late main sequence star companion in close orbits around sdB stars.
\end{abstract}

\maketitle

%%%%%%%%%%%%%%%%%%%%%%%%%%%%%%%%%%%%%%%%%%%%
%% MAINMATTER
%%%%%%%%%%%%%%%%%%%%%%%%%%%%%%%%%%%%%%%%%%%%

\section{Introduction}
The formation of sdBs is still unclear as it requires an extraordinarily high mass loss on the red-giant branch. About half of the sdB stars are found in close binaries with periods down to $\sim 0.1$\,d (e.g.\ Maxted et al.\ 2001). Performing binary evolution studies Han et al.\ (2003) found that common-envelope (CE) evolution, resulting from dynamically unstable mass transfer near the tip of the first RGB, should produce such short-period binaries ($P=0.1-10\,{\rm d}$). The most probable mass for such sdBs is 0.47 $M_{\odot}$, which is called the canonical mass.\\
But there were also sdBs found with masses too low to sustain helium burning in their core. They are believed to form by an extreme mass loss on the red giant branch before they can ignite helium. Such stars are called post RGB stars. They evolve directly to white dwarfs. Driebe et al.\ (1998) analysed the evolutionary tracks of such stars and found that they cross the area of the EHB where the sdBs lie.\\
\begin{figure}[!t]
%		\includegraphics[width=0.33\textwidth]{lcasasb}
%		\hfill%\hspace{0.25cm}
%  		\includegraphics[width=0.33\textwidth]{lcasasv}
%  		\hfill%\hspace{0.25cm}
%  		\includegraphics[width=0.33\textwidth]{lcasasi}
%  		\caption{B, V and I ASAS lightcurves of ASAS\,10232.}
\parbox{\textwidth}{% 
\begin{center}
		\includegraphics[width=0.48\textwidth]{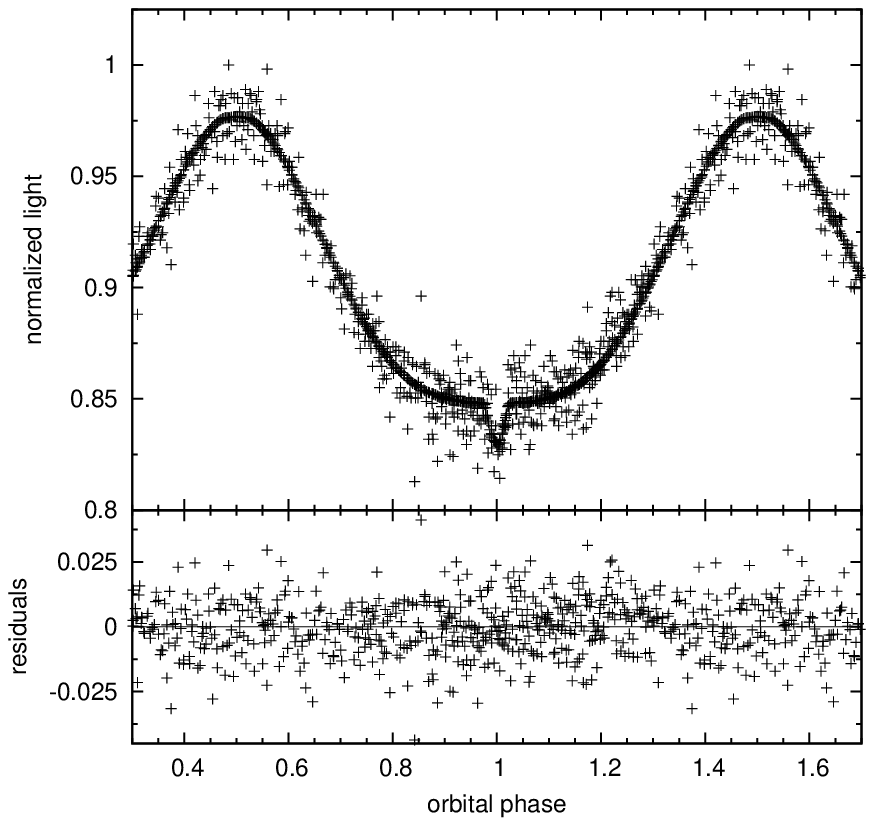}
%\\%		\hfill%\hspace{0.25cm}
  		\includegraphics[width=0.48\textwidth]{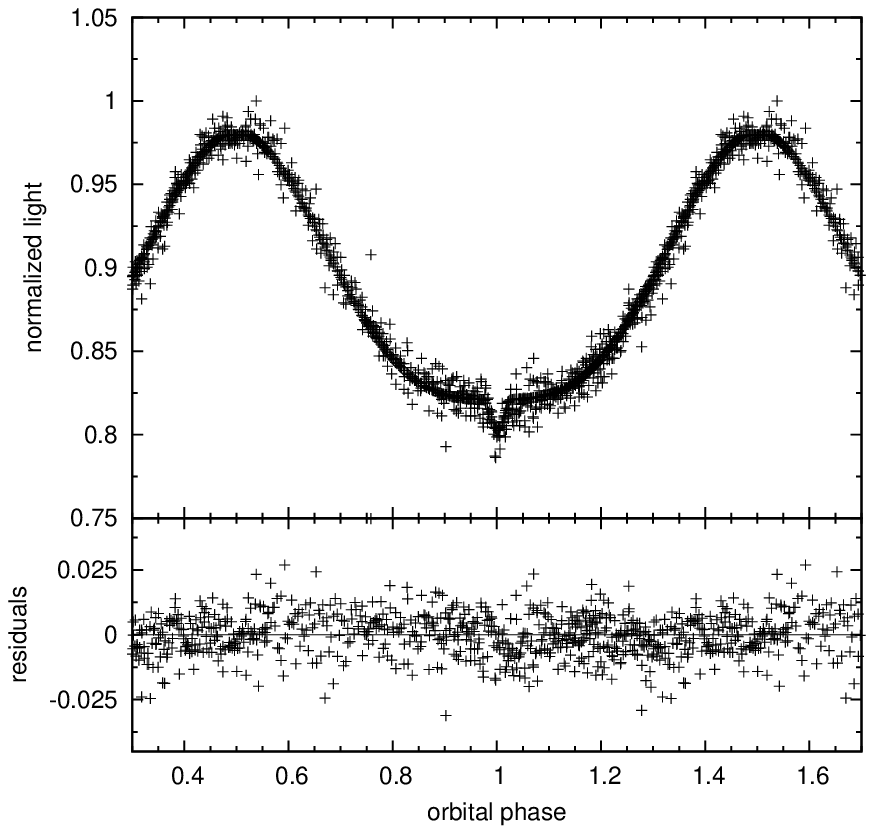}
\\%  		\hfill%\hspace{0.25cm}
  		\includegraphics[width=0.48\textwidth]{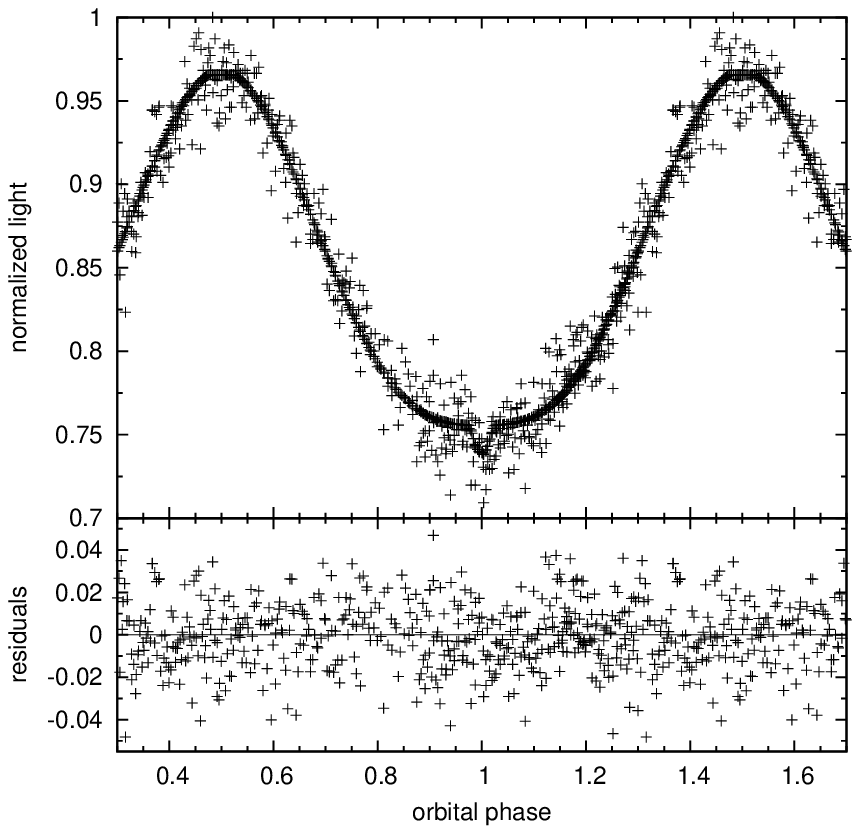}
\end{center}
}
  		\caption{B, V and I ASAS light curves of ASAS\,10232.} 
\end{figure}
Amongst the close binaries eclipsing binaries are of high value because it is possible to derive the mass of the sdB as well as the mass and nature of the companion from a combined analysis of time resolved spectra and light curve. The distribution of sdB masses is one of the most important constraints to identify the formation channel among several possibilities. HW Vir type binaries are a group of detached binary systems that consist of sdBs and late M star companions with periods of about $2-3\,{\rm hr}$. Because of the short possible periods and the similar size of both components such systems have high probability to be eclipsing. Until now only nine such systems are known.\\
Soker (1998) proposed the idea that planetary or brown dwarf companions could cause the mass loss necessary to form an sdB. Substellar objects with masses higher than $>10\,M_{\rm J}$ were predicted to survive the common-envelope phase and end up in a close orbit around the stellar remnant, while planets with lower masses would entirely evaporate.\\
Following this Nelemans and Tauris (1998) suggested a scenario to form low mass white dwarfs accompanied by a substellar object. They calculated that the survival of the companion also depends on the orbital separation (M $> 25 M_J$, a $> 0.4 R_{\odot}$).
\begin{figure}[!Ht]
		\includegraphics[width=0.49\textwidth]{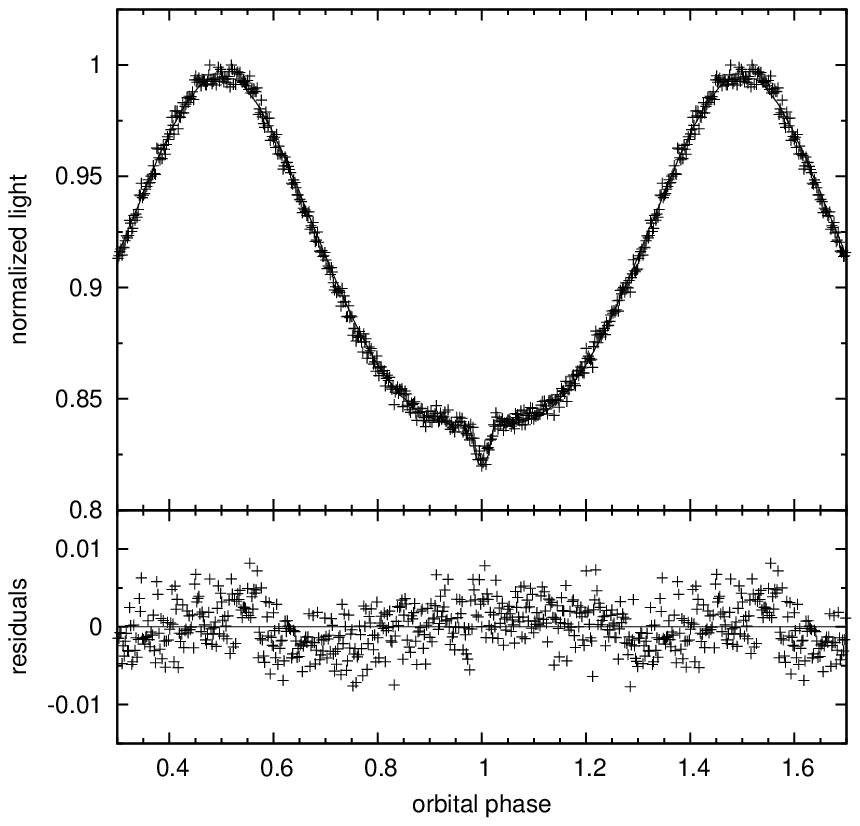}
		\hfill%\hspace{0.5cm}
  		\includegraphics[width=0.49\textwidth]{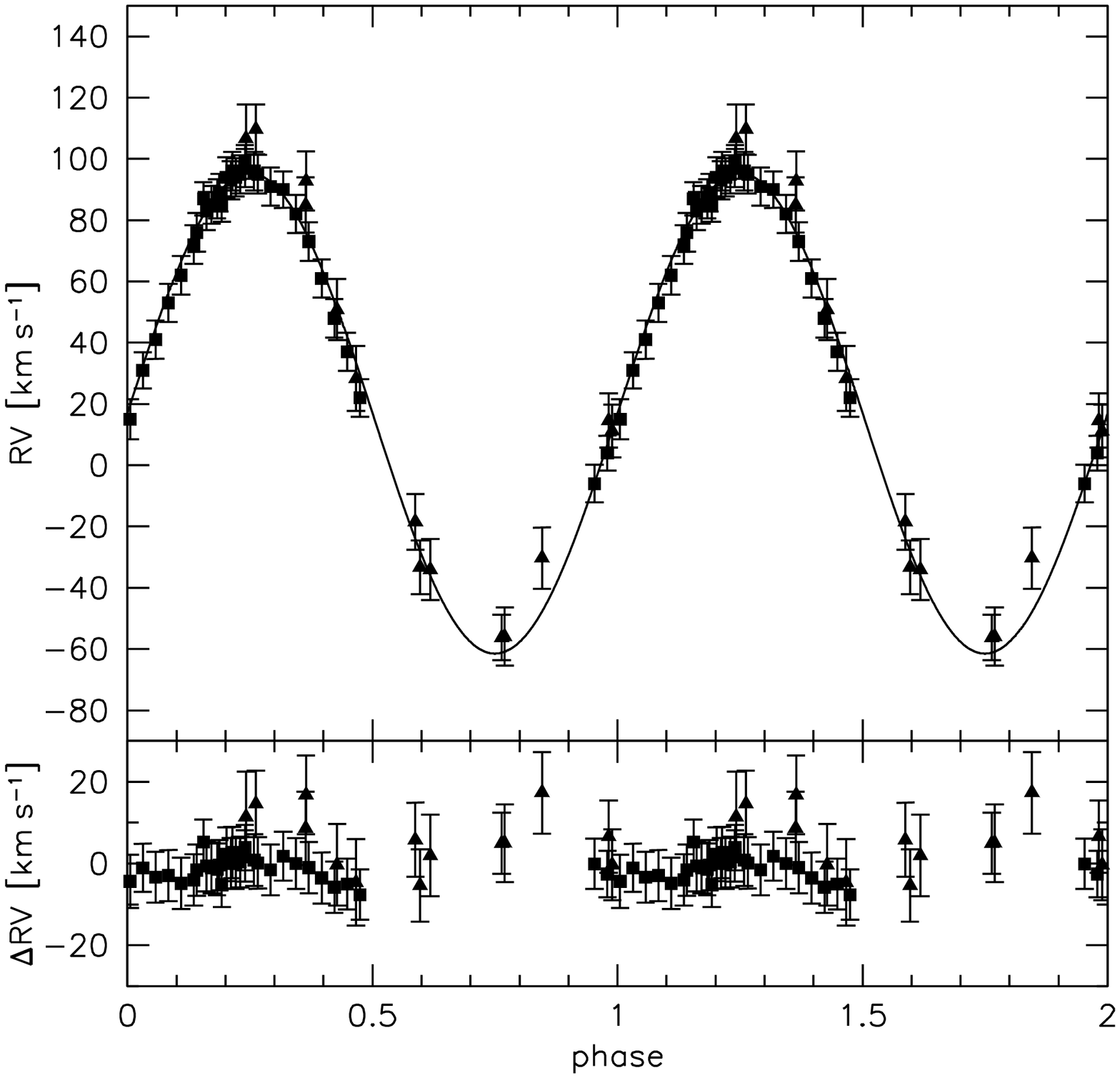}
  		\caption{Radial velocity curve and SuperWASP light curve ASAS 10232.} 
\end{figure} 
\section{Analysis}
ASAS\,102322-3737.0 (ASAS\,10232 for short) was discovered by inspection of light curves in the B,V,\,I bands obtained from the ASAS survey (Pojmanski 2002, see Fig.~1). A white light curve from SuperWASP (Pollacco et al.\ 2006, see Fig.~2) revealed a grazing eclipse. 46 spectra of ASAS\,102322 have been taken with ESO\,NTT/EMMI and Gemini/GMOS.\\
SDSS\,J082053.53+000843.4 (J082053 for short) was discovered in the course of the MUCHFUSS project (see Geier et al.\ these proceedings). After detecting radial velocity variations with short orbital period, we obtained a light curve in the R band with the Mercator telescope. 25 spectra have been taken with SDSS, ESO\,NTT/EFOSC2 and SOAR/Goodman.\\ 
All light curves were analysed with the MORO code (Drechsel et al.\ 1995). We used Wilson-Devinney mode 2, which poses no restrictions to the system configuration and links the luminosity and the temperature of the second component by means of the Planck law. The gravity darkening exponent and the linear limb darkening coefficients were fixed at literature values. The temperatures of the hot components were taken from the spectral analysis. For the albedos of the companion values > 1 were necessary to model the reflection effect. This can be explained with processes in the stellar atmosphere that cause a spectral redistribution of the irradiated energy in the wavelengths. The radiation pressure coefficient for the primary star, the inclination, the effective temperature of the companion and the Roche potentials were adjusted. A grid with different fixed mass ratios and different start parameters was calculated. In the case of ASAS\,10232 all four light curves were analysed simultaneously.\\
All spectra have medium resolution ($R\simeq1200-3400$). Atmospheric parameters have been determined by fitting LTE model spectra with solar metallicity (Heber et al.\ 2000) to the hydrogen Balmer and helium lines. The radial velocities (RV) were measured by fitting Gaussians, Lorentzians and polynomials to the hydrogen Balmer lines as well as helium lines. Assuming circular orbits sine curves were fitted to the RV data points and the orbital parameters were derived.
\section{ASAS 10232}
The light curve shows a very prominent reflection effect as well as a grazing primary eclipse (see Fig.~2). No secondary eclipse can be seen. Therefore the parameters are less well constrained than in the case of J082053 (next sect.). The solution with the smallest standard deviation is shown in Table~1. The Super WASP and ASAS light curves and the radial velocity curve together with the best fit are given in Figs.~1 and 2. With a period of 0.139 d ASAS 10232 has one of the longest periods known for HW Virginis systems.% \\

The sdB mass of $0.46\,M_{\rm \odot}$ is consistent with theory. The surface gravity derived from the spectra is in agreement with the one from photometry and the radius of the $0.15\,M_{\rm \odot}$ companion is in conformity with theoretical mass-radius relations for main sequence stars. Although the sdB is more massive, the companion is bigger and slightly distorted; {\it ASAS\,10232 is therefore most likely a typical HW\,Vir system seen at lower inclination.}\\
%\subsection{Radial velocity curve}
%\begin{figure}[h]
  	%	
  	%\end{figure}
%
\begin{figure}[t]
  		\includegraphics[width=0.495\textwidth]{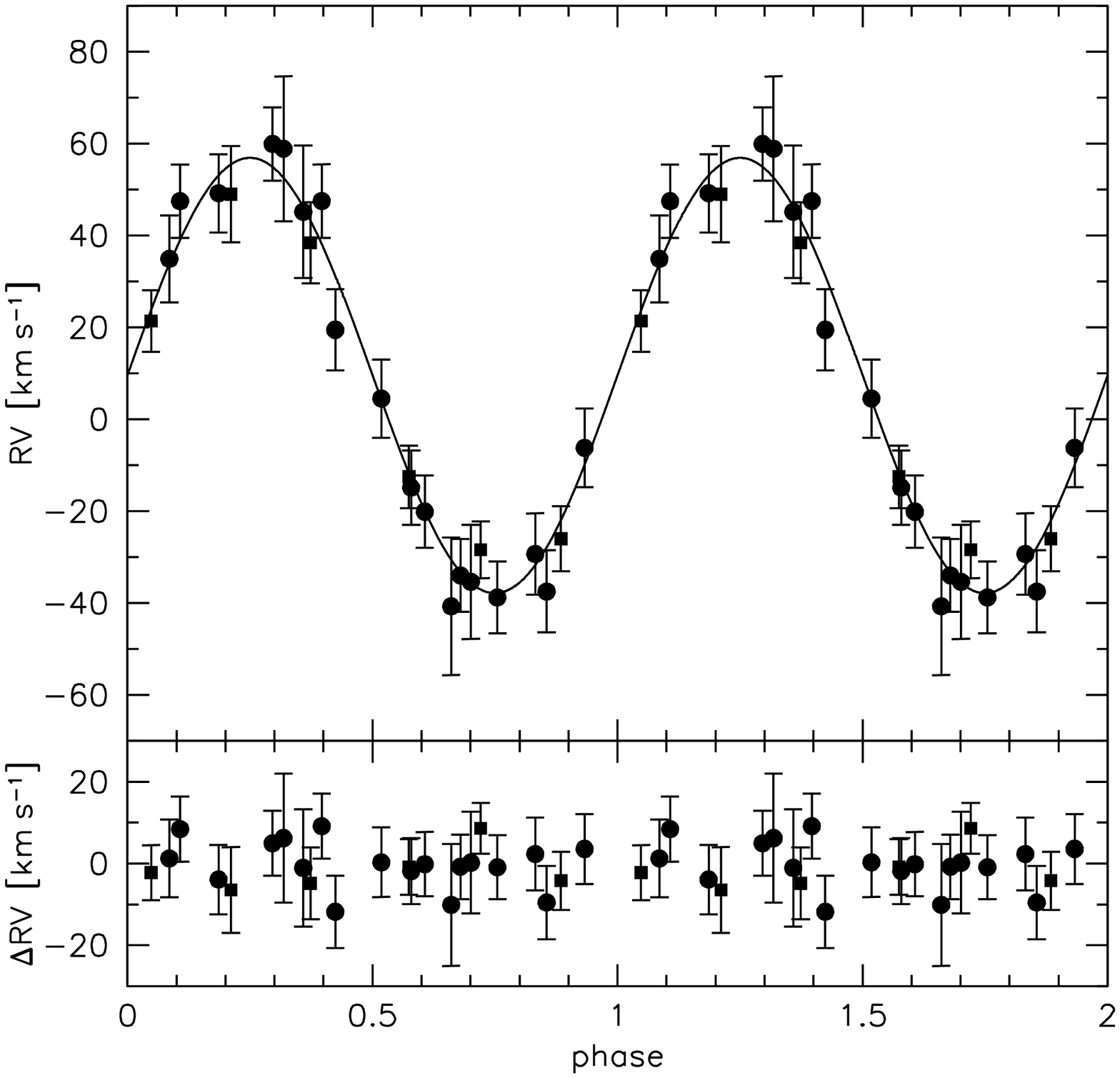}
  		\hfill%\hspace{0.5cm}
  		\includegraphics[width=0.495\textwidth]{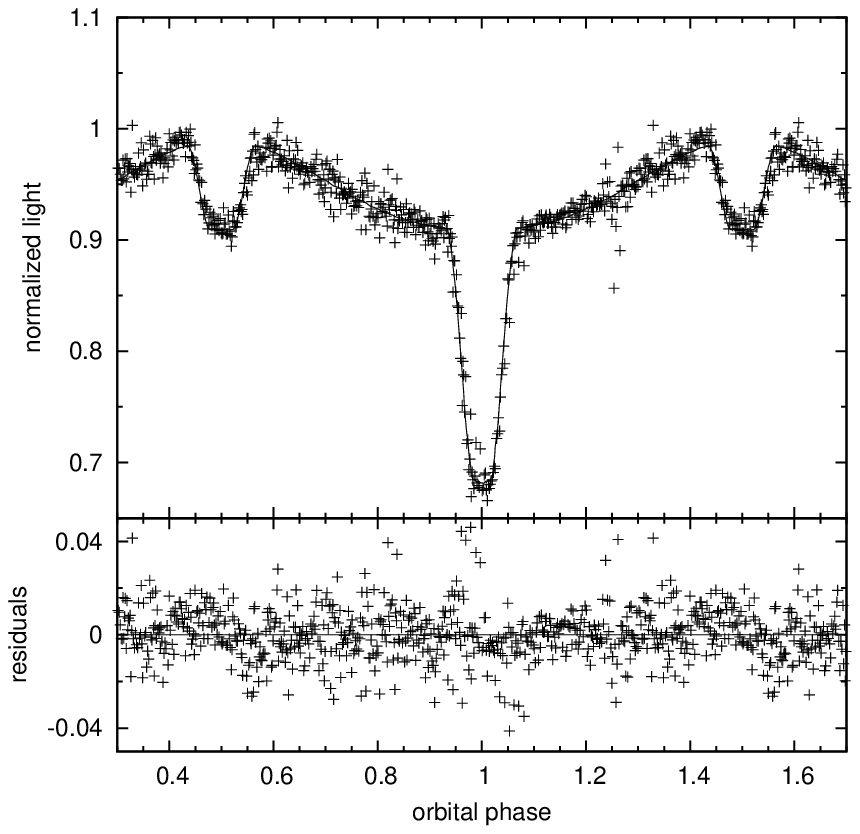}
  	\caption{Radial velocity curve and Mercator light curve of J082053.}
\end{figure}
\section{J082053}
The light curve shows total eclipses (see Fig.~3). The apparent reflection effect indicates a HW Virginis System. 
We performed an analysis of the RV and the light curve in the same way as described for ASAS10232. The results are summarized in Table~2. As can be seen in Fig.~4, solutions over a huge mass range for the sdB are possible. 
\begin{table}[!Ht]	
		\begin{tabular}{llll}
		\noalign{\smallskip}
		\hline
		Spectroscopy&&& \\
		\noalign{\smallskip}
		\hline
		\noalign{\smallskip}
		Effective temperature & $T_{\rm eff,sdB}$ & [K] & $28\,400\pm500$\\
		Surface gravity & $\log{g}_{\rm sdB,spec}$           & & $5.56\pm0.05$\\
		Orbital period & $P$ & [d] & $0.13927$\\
		RV semi-amplitude & $K$ & [${\rm km~s^{-1}}$] & $81.0\pm3.0$\\
		\noalign{\smallskip}
		\hline
		\noalign{\smallskip}
		Light curve parameter&&&\\
		\noalign{\smallskip}
		\hline
		Orbital inclination & $i$ & [$^{\rm \circ}$] & $65.858\pm 0.690$ \\
		Mass ratio & $q=M_{2}/M_{1}$ & & $0.34$\\
		Effective temperature & $T_{\rm eff,comp}$ & [K] & $3332 \pm 553$\\
		%Albedo & $A_1$ & & $0.944\pm0.027$\\
      	%	   & $A_2$ & & $1.207\pm0.131$\\
		Roche radii & $r_1$ & [$a$] & $0.18540\pm 0.00888$\\
			   & $r_2$ & [$a$] & $0.26627\pm 0.01213$\\
		\noalign{\smallskip}
		\hline
		\noalign{\smallskip}
		{\bf Derived parameter}&&&\\
		\noalign{\smallskip}
		\hline
		Subdwarf mass & $M_{\rm sdB}$ & [$M_{\rm \odot}$] & $0.461\pm0.051$\\
		Companion mass & $M_{\rm comp}$ & [$M_{\rm \odot}$] & $0.157\pm0.017$\\
		Separation & $a$ & [$R_{\rm \odot}$] & $0.963\pm 0.036$\\
		Surface gravity& $\log{g}_{\rm sdB,phot}$ & & $5.60\pm0.02$\\
		\noalign{\smallskip}
		\hline
		\end{tabular}
		\caption{Parameters of ASAS 10232}
		\label{param_asas}
		\end{table}
\begin{table}[!Ht] 
	\begin{tabular}{llll}
		\noalign{\smallskip}
		\hline
		\noalign{\smallskip}
		Spectroscopy && best solution&canonical mass \\
		\noalign{\smallskip}
		\hline
		\noalign{\smallskip}
		$T_{\rm eff,sdB}$ & [K] & \multicolumn{2}{c}{$26\,100\pm900$}\\
		$\log{g}_{\rm sdB,spec}$ & &\multicolumn{2}{c}{ $5.48\pm0.07$}\\
		$P$ & [d] & \multicolumn{2}{c}{$0.09603\pm0.001$}\\
		$K$ & [${\rm km~s^{-1}}$] & \multicolumn{2}{c}{ $47.4\pm2.0$}\\
		\noalign{\smallskip}
		\hline
		\noalign{\smallskip}
		Light curve parameter&&&\\
		\noalign{\smallskip}
		\hline
		$i$ & [$^{\rm \circ}$] & $85.870\pm0.164$ & $85.829\pm0.193$ \\
		$q=M_{2}/M_{1}$ & & $0.181$ & $0.1438$\\
		$T_{\rm eff,comp}$ & [K]& $2880\pm202$ & $2419\pm224$\\
		%$A_1$ (fixed) & & $1.0$ & $1.0$ \\
		%$A_2$ & & $1.11\pm0.05$ & $1.09\pm0.04$\\
		$r_1$ & [$a$] & $0.2804\pm0.0022$ & $0.2820\pm0.0022$\\
		$r_2$ & [$a$] & $0.1375\pm0.0010$ & $0.1386\pm0.0011$\\
		$\sigma_{\rm fit}$ & & $0.01233$ & $0.01235$\\
		\noalign{\smallskip}
		\hline
		\noalign{\smallskip}
		{\bf Derived parameter}&&&\\
		\noalign{\smallskip}
		\hline
		$M_{\rm sdB}$ & [$M_{\rm \odot}$] & $0.251\pm0.032$  & $0.47\pm0.06$\\
		$M_{\rm comp}$ & [$M_{\rm \odot}$] & $0.045\pm0.006$ & $0.068\pm0.009$\\
		$a$ & [$R_{\rm \odot}$]& $0.588\pm0.026$ & $0.717\pm0.031$\\
		$\log{g}_{\rm sdB,phot}$ & & $5.41\pm0.02 $ &$5.50\pm0.02$\\
		\noalign{\smallskip}
		\hline
	\end{tabular}
	\caption{Parameters of J082053}
\end{table} 
\begin{figure}[t]
  		\includegraphics[width=0.78\textwidth]{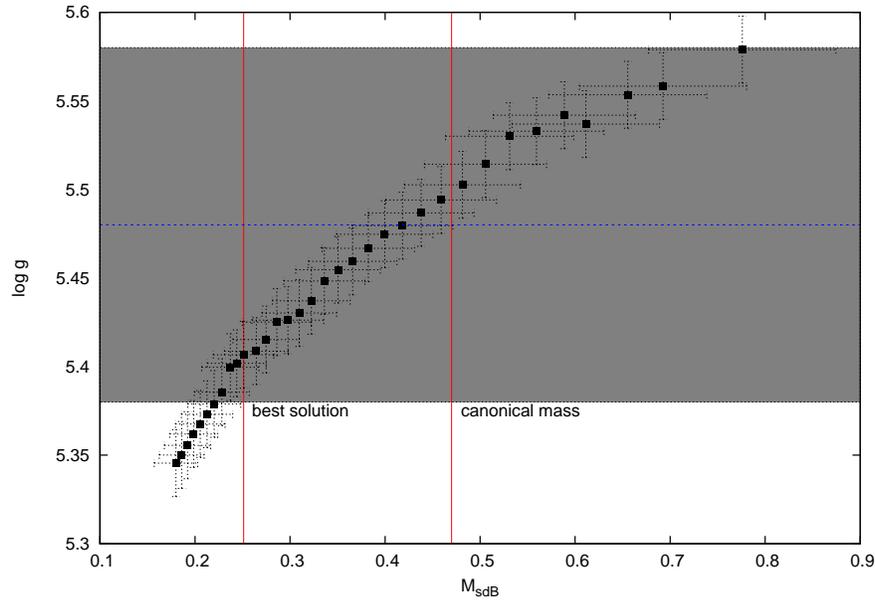}
  	\caption{$M_{\rm sdB}-\log{g}$ diagram for the different solutions, the grey area represents the possible values for $\log{g}$ from the spectroscopic analysis.}
\end{figure} 
\begin{figure}[t]
  		\includegraphics[width=0.82\textwidth]{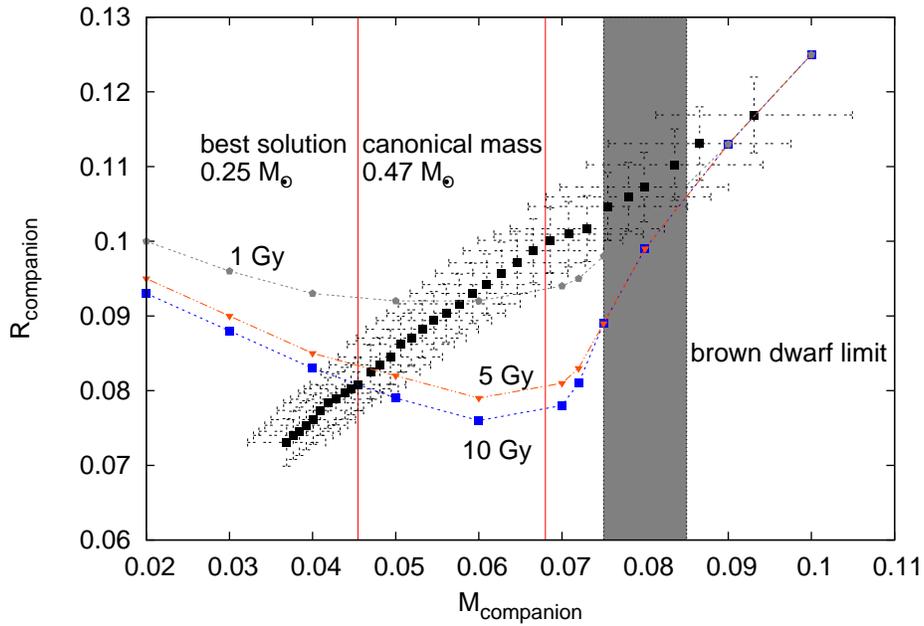}
  	\caption{Comparison of measured mass-radius relation for the companion with theoretical mass radius relations for brown dwarfs for different ages (Baraffe et al.\ 2003).}
\end{figure}
Therefore we consider a solution for the canonical sdB mass ($0.47\,M_{\rm \odot}$), which results in a substellar companion mass of $0.068 M_{\rm \odot}$. Higher masses are not predicted by theoretical binary evolution studies (Han et al.\ 2003) nor supported by mass estimates from asteroseismology and the analysis of other HW Virginis Systems. 
\begin{figure}[h]
  		\includegraphics[width=0.495\textwidth]{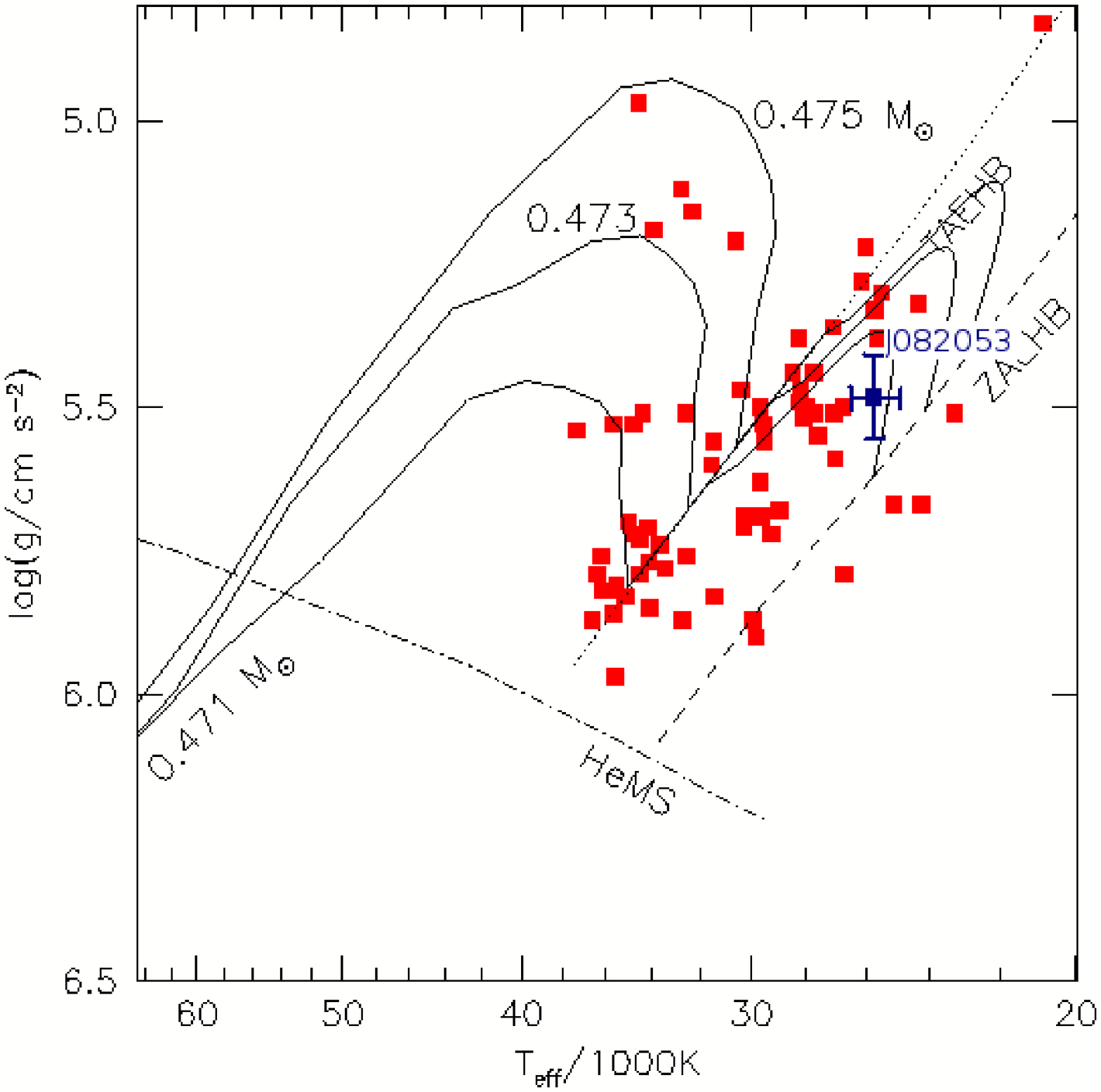}
  		\hfill%\hspace{0.5cm}
  		\includegraphics[width=0.495\textwidth]{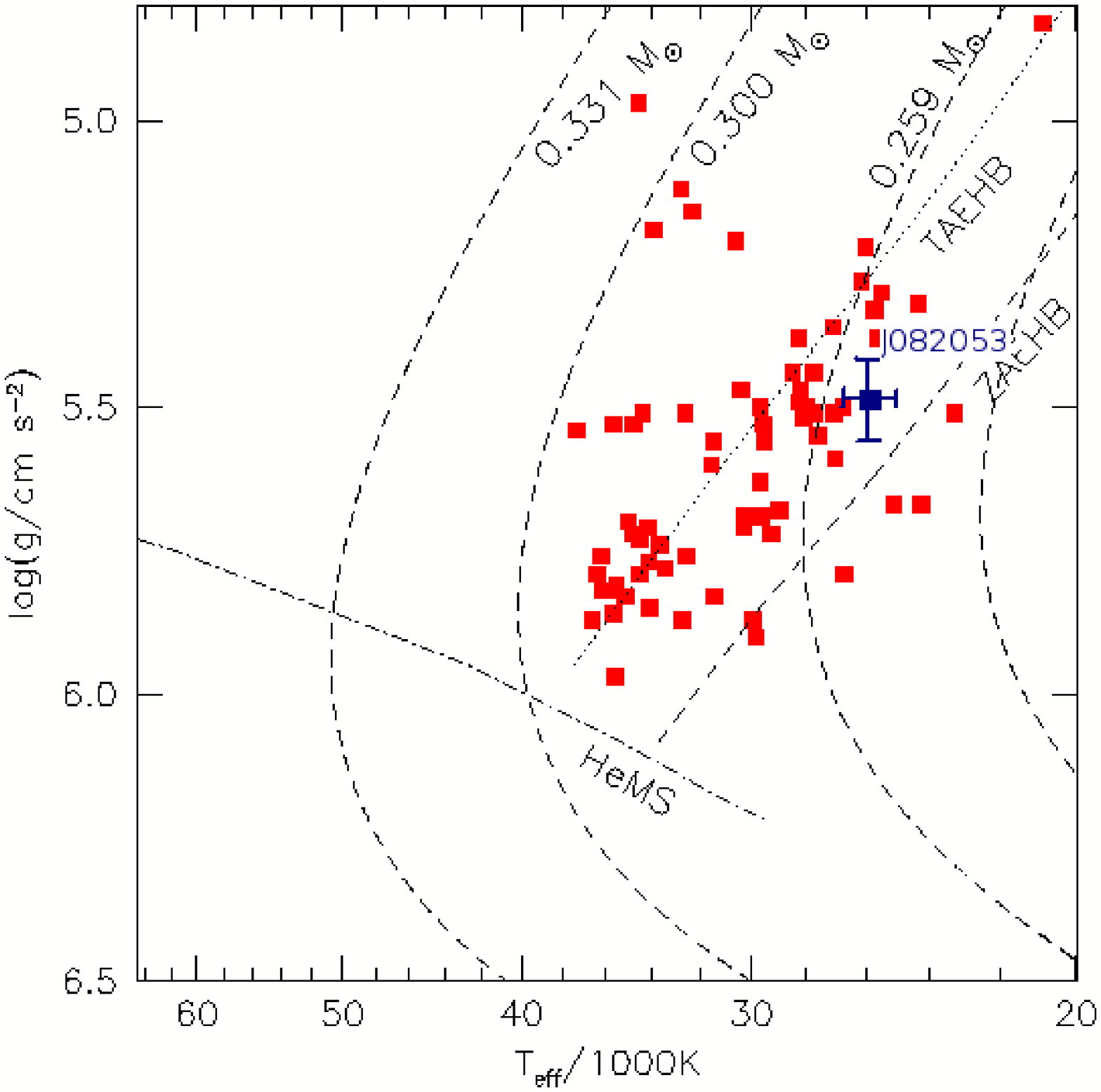}
  	\caption{comparison of position of J082053 in the $\log{g}-T_{\rm eff}$ diagram with evolutionary tracks of helium burning cores (left, from Dorman et al.\ 1993) and tracks from non helium burning cores (right, from Driebe et al.\ 1998).}
\end{figure}
However, the companion's radius is larger by 20\% than predicted by the mass-radius relation for brown dwarfs as it can be seen in Fig.~5. As the sdB star has probably evolved from a progenitor of $\approx 1 M_{\odot}$ on the main sequence, we expect the system to be old. Hence we compare with the mass-radius relations for 5 Gyrs and 10 Gyrs. The bigger radius may be due to the inflation of the companion by the irradiation by the sdB primary. This effect was not measured yet for sdBs and cannot be evaluated. However the sdB progenitor  could have been more massive ($1 M_{\odot}-2 M_{\odot}$) and therefore younger. In this case the sdB mass may be lower ($0.3 M_{\odot}-0.4 M_{\odot}$) because helium burning starts under non-degenerate conditions (see Hu et al.\ 2009). The companion mass would then be ($\approx 0.06 M_{\odot}$), perfectly consistent with the 1 Gyr mass-radius relation for brown dwarfs.
We also examined another solution, actually the solution with the smallest $\chi^2$. The radius of the companion is consistent with the BD mass-radius relation (see Fig.~5). It gives an sdB mass of $0.25 M_{\rm \odot}$ and a companion mass of $0.045 M_{\rm \odot}$. The low sdB mass implies that the star does not burn helium in the core but evolves into a helium white dwarf (Driebe et al.\ 1998), i.e. it is a post-RGB star. 
In Fig.~6 a comparison of the position of J082053 in the $\log{g}-T_{\rm eff}$ diagram with the evolutionary tracks of either EHB or post RGB stars is shown. The atmospheric parameters of the object are consistent with an sdB and a mass of $0.47 M_{\rm \odot}$ or a post RGB star with a mass of $0.25 M_{\rm \odot}$. Hence this two solutions cannot be distinguished. We regard it more likely that it is an EHB star rather than a post-RGB star since the latter seem to be rare. The mass of the companion is below the limit for stable hydrogen burning for both solutions. {\it Therefore the companion is most likely a brown dwarf -- the first one unambiguously identified in an sdB binary.}\\ If the star is a post RGB this system will evolve into a BD-WD system such as WD~0137$-$349 (see Burleigh these proceedings) as soon as the temperature of the primary star cooled down to 15\,000 K.
\section{outlook}
Moreover, we discovered a new HW Virginis System in a run on the Calar Alto Observatory with BUSCA in June. There was not enough time to analyse it yet. The light curve is shown in Fig.~7. The period of the system is 0.075 d. This is the shortest period known for a HW Virginis system so far. It is a candidate for the second sdB-brown dwarf system. 
 \begin{figure}[t]
  		\includegraphics[height=0.95\textwidth, angle=-90]{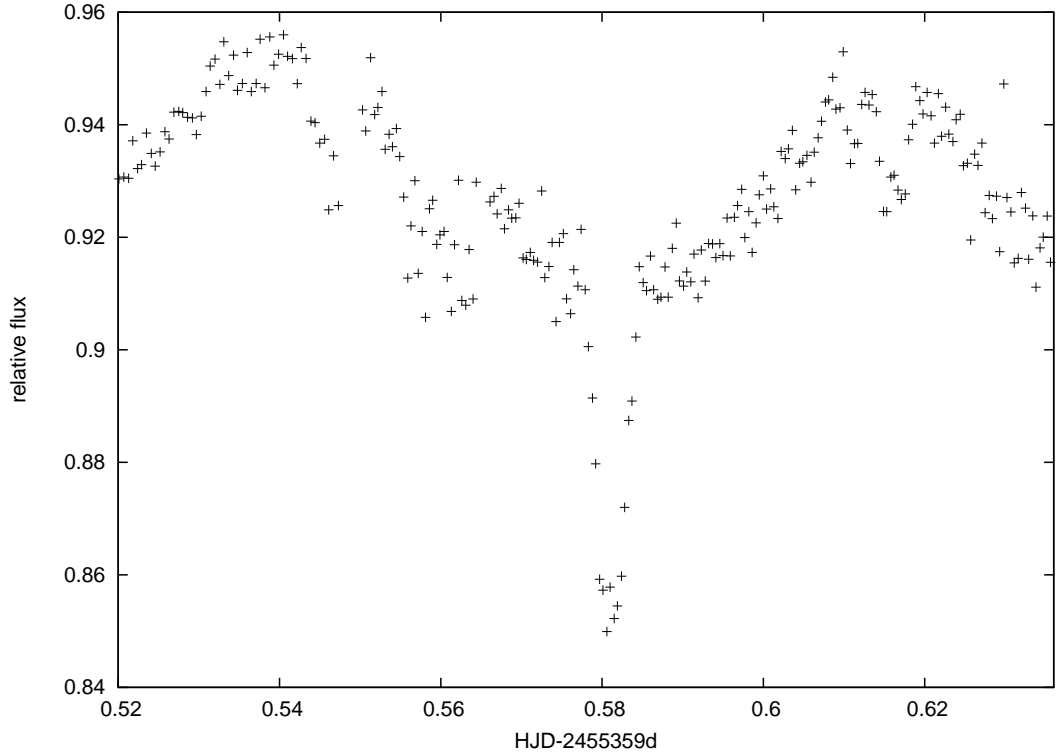}
  	\caption{$U_B$ light curve of J162256+473051, obtained with BUSCA in June 2010.}
\end{figure}

%\subsection{Radial velocity curve}	

%%%%%%%%%%%%%%%%%%%%%%%%%%%%%%%%%%%%%%%%%%%%%%%%
%% BACKMATTER
%%%%%%%%%%%%%%%%%%%%%%%%%%%%%%%%%%%%%%%%%%%%%%%%

%\begin{theacknowledgments}
%  Infandum, regina, iubes renovare dolorem, Troianas ut opes et
%  lamentabile regnum cruerint Danai; quaeque ipse miserrima vidi, et
%  quorum pars magna fui. Quis talia fando Myrmidonum Dolopumve aut duri
%  miles Ulixi temperet a lacrimis?
%\end{theacknowledgments}




\begin{thebibliography}{6}
\bibitem{baraffe:2003}
Baraffe, I., Chabrier, G.; Barman, T.~S., Allard, F., Hauschildt, P.~H., \emph{A\&A}, \textbf{402}, 701 (2003).
\bibitem[(Burleigh et al.\ 2011)]{Burleigh}
Burleigh, M.~R., et al.,
''White Dwarf - Brown Dwarf Binaries'',
in \emph{Planetary Systems beyond the Main Sequence},
edited by S.~Schuh, H.~Drechsel and U.~Heber, AIP Conference Proceedings,
American Institute of Physics, New York (these proceedings).
\bibitem{Drechsel:1995}
Drechsel, H., Haas, S., Lorenz, R., Gayler, S., \emph{A\&A}, \textbf{294}, 723 (1995).
\bibitem{Driebe1998}
Driebe, T., Schoenberner, D., Bloecker, T., Herwig, F.~T., \emph{A\&A}, \textbf{339}, 123 (1998).
\bibitem{dorman:1993}
Dorman, B., Rood, R.~T., O'Connell, R.~W., \emph{APJ},  \textbf{419}, 596 (1993). 
\bibitem[(Geier et al.\ 2011)]{Geier}
Geier, S., Heber, U., Tillich, A., et al.,
''Substellar Companions and the Formation of Hot Subdwarf Stars'',
in \emph{Planetary Systems beyond the Main Sequence},
edited by S.~Schuh, H.~Drechsel and U.~Heber, AIP Conference Proceedings,
American Institute of Physics, New York (these proceedings).
\bibitem{Han:2003}
Han, Z., Podsiadlowski, P., Maxted, P.~F.~L., \& Marsh, T.~R., \emph{MNRAS}, \textbf{341}, 669 (2003).
\bibitem{Heber:2000}
{{Heber}, U., {Reid}, I.~N., {Werner}, K.}, \emph{A\&A}, \textbf{363}, 198 (2000).
\bibitem{Hu:209}
{Hu}, H., {Nelemans}, G., {Aerts}, C., {Dupret}, {M.-A.}, \emph{A\&A}, \textbf{508}, 869 (2009).
\bibitem{maxted, 2001}
{Maxted}, P.~F.~L., {Heber}, U., {Marsh}, T.~R., {North}, R.~C., \emph{MNRAS}, \textbf{333}, 231 (2001).
\bibitem{Nelemans and Tauris, 1998}
Nelemans, G. and Tauris, T.~M., \emph{A\&A}, \textbf{335}, 85 (1998).
\bibitem{pojmanski, 2002}
Pojmanski, G., \emph{Acta Astronomica}, \textbf{52}, 397 (2002).
\bibitem{pollaco, 2006}
Pollacco, D., \emph{PASP}, \textbf{118}, 1407 (2006).
\bibitem{Soker}
Soker, N., 1998, \emph{AJ}, \textbf{116}, 1308 (1998).
\end{thebibliography}
\end{document}